\documentclass[]{spie}  
\usepackage[dvips]{graphicx}

\newcommand{\teff}{$T_{\rm eff}$}
\newcommand{\msun}{$M_\odot$}
\newcommand{\lessim}{$^{<}_{\sim}$}

\title{Data reduction and astrometry strategies for wide-field images:
an application to the Capodimonte Deep Field} 

\author{Alcal\'a J.M.\supit{a}, Radovich M.\supit{a}, Silvotti R.\supit{a},
Pannella M.\supit{a}, Arnaboldi M.\supit{a},\\ Capaccioli M.\supit{a,b}, Longo G.\supit{b}
\skiplinehalf
\supit{a}INAF-Osservatorio Astronomico di Capodimonte, Via Moiariello 16
I-80131, Naples, Italy \\
\supit{b} University Federico II, Naples, Italy
}

\authorinfo{Further author information: 
(Send correspondence to J.M. Alcal\'a)\\
J.M. Alcal\'a: E-mail: jmae@na.astro.it, Telephone: +39 0815575 479\\  
M. Radovich: E-mail: radovich@na.astro.it, Telephone: +39 0815575 446, 
Address: Via Moiariello 16, I-80131, Naples, Italy}

\begin{document} 
  \maketitle 

\begin{abstract}
The Capodimonte Deep Field (OACDF) is a multi-colour imaging survey on two 
0.5x0.5 square degree fields performed in the BVRI bands and in six 
medium-band filters (700 - 900 nm) with the Wide Field Imager (WFI) at the 
ESO 2.2 m telescope at La Silla, Chile. In this contribution the adopted 
strategies for the  OACDF data reduction are discussed. Preliminary
scientific results of the survey are also presented.

\end{abstract}


\keywords{CCD data reduction, wide field imaging, surveys}


\section{Introduction}
In view of the arrival of the VLT Survey Telescope (VST) (a cooperation 
of ESO, the Capodimonte Astronomical Observatory-OAC-Naples, Italy, and 
the European Consortium Omegacam, for the design, realization, installation, 
and operation at ESO Paranal Observatory in Chile of a 2.6 m aperture, 
1 degree x 1 degree wide field imaging facility in the spectral range 
from UV to I bands), the OAC started a pilot project, called the Capodimonte 
Deep Field (OACDF), consisting in a multi-colour imaging survey using the 
Wide Field Imager (WFI, Baade et al. 1998 ) at the ESO 2.2~m telescope at 
La Silla, Chile. The VST will be equipped with OmegaCam, a 16k X 16k array 
of 32 CCDs, which will cover 1 square degree. 
The main goal of the OACDF is to provide a large photometric database, mainly 
oriented to extragalactic studies (quasars, high-redshift galaxies, galaxy 
counts, lensing, etc.), that can be used also for stellar and planetary 
research (galactic halo population, peculiar objects like brown dwarfs and 
cool white dwarfs, Kuiper-Belt objects). Another goal is to gain insight into 
the handling and processing of data coming from a wide field imager, similar 
to the one which will be installed at VST. 

In this contribution we discuss technical aspects of the OACDF data reduction 
and present some preliminary scientific results. We focus on the following 
topics: super flat-fielding and defringing of the mosaics, astrometry and  
photometric calibration. Such topics will be of primary importance for the 
processing of the VLT Survey Telescope (VST) data, in the framework of the 
European consortium ASTRO-WISE (see http://www.ASTRO-WISE.org). Some of the 
problems encountered are intrinsic to the ESO WFI. Hence, our tests might 
contribute to a better characterisation of this instrument. In addition, we 
report on some preliminary scientific results of the OACDF project. The depth 
of the OACDF (25.1 mag in the R band) allows to achieve the foreseen 
scientific goals: i) the search for rare/peculiar objects, AGNs and 
high-redshift QSO's (z$>$3); ii) the search for intermediate-redshift 
early-type galaxies to be used as tracers of galaxy evolution and 
iii) the search for galaxy clusters to be used as targets for spectroscopic 
follow-up's at larger telescopes.


\section{Observations} 
The observations for the OACDF project were performed in three 
different periods (18--22 April  1999, 7--12 March 2000 and 26--30 
April 2000) using the WFI mosaic camera attached to the ESO 2.2m 
telescope at La Silla, Chile.
This camera consists of eight 2k$\times$4k CCDs that constitute a 
8k$\times$8k array. The scale is $0.238''$/pix.
Some 100 Gbyte of raw data were collected in each one of the three 
observing runs. The first and third run were photometric, while the
second one was partially non-photometric.

The observational strategy was to perform a 1~deg$^2$ shallow 
survey and a 0.5~deg$^2$ deep survey.
The shallow survey  was performed in the B,V,R and I broad-band filters.
Four adjacent $30'\times30'$ fields, covering a $1^{\circ} \times 1^{\circ}$
field in the sky, were observed for the shallow survey. We call these 
fields OACDF1, OACDF2, OACDF3 and OACDF4.
The OACDF2 and OACDF4 were chosen for the 0.5~deg$^2$ deep 
survey. 
The deep survey was performed in the B,V,R broad-bands and in six 
medium-band filters (700 - 900 nm).  

Several standard fields, selected from the Landolt (1992) E-regions
were observed in order to transform the BVRI instrumental magnitudes 
to the standard Johnson/Kron-Cousins system. Likewise, several 
spectrophotometric standard stars were observed in the intermediate-band
filters for absolute flux calibration purposes. 

Being obtained with a mosaic imager, the WFI data are multi-extension
fits files (MEFs). Each extension corresponds to one of the eight 
CCDs of the mosaic. The physical separation of each one of the CCDs 
with respect to each other is about 100$\mu$. A sequence of ditherings 
following a rhombi-like pattern were performed in order to cover the 
CCD gaps. For the shallow survey a sequence of 5 ditherings was done, 
while at least 8 ditherings were  done for the deep survey.

\section{Data reduction} 
A two-processor (with 500 Mhz each) DS20 machine with hundred
Gbyte of hard disk was used for the data reduction. The {\it mscred} task 
under IRAF\footnote{IRAF is distributed by the National Optical Astronomy 
Observatories, which is operated by the association of the universities 
for research in astronomy, Inc., under contract with the National Science 
Fundation.} was used to perform the data reduction. The reduction steps
are described next.

\subsection{Bias and dark}
 A mean Bias was created nightly. The bias subtraction was performed 
 extension by extension. Several sixty-minute dark exposures were used 
 to derive average dark frames for each night. 
 No significant contribution from the mean dark to the noise was 
 found.  

\subsection{Flat-fielding}
 A nightly super-flat was created as the result of the median of
 all the science frames taken in the same filter. 
 We define  $supFlat= \langle science~frames \rangle$.
 By means of a rejection algorithm, all astronomical sources over 
 the background level,
 as well as  cosmic particle heats and bad columns are normally removed
 in the resulting $supFlat$ image, because of the dithering pattern.

\begin{figure}  
\vspace{7cm}
\includegraphics{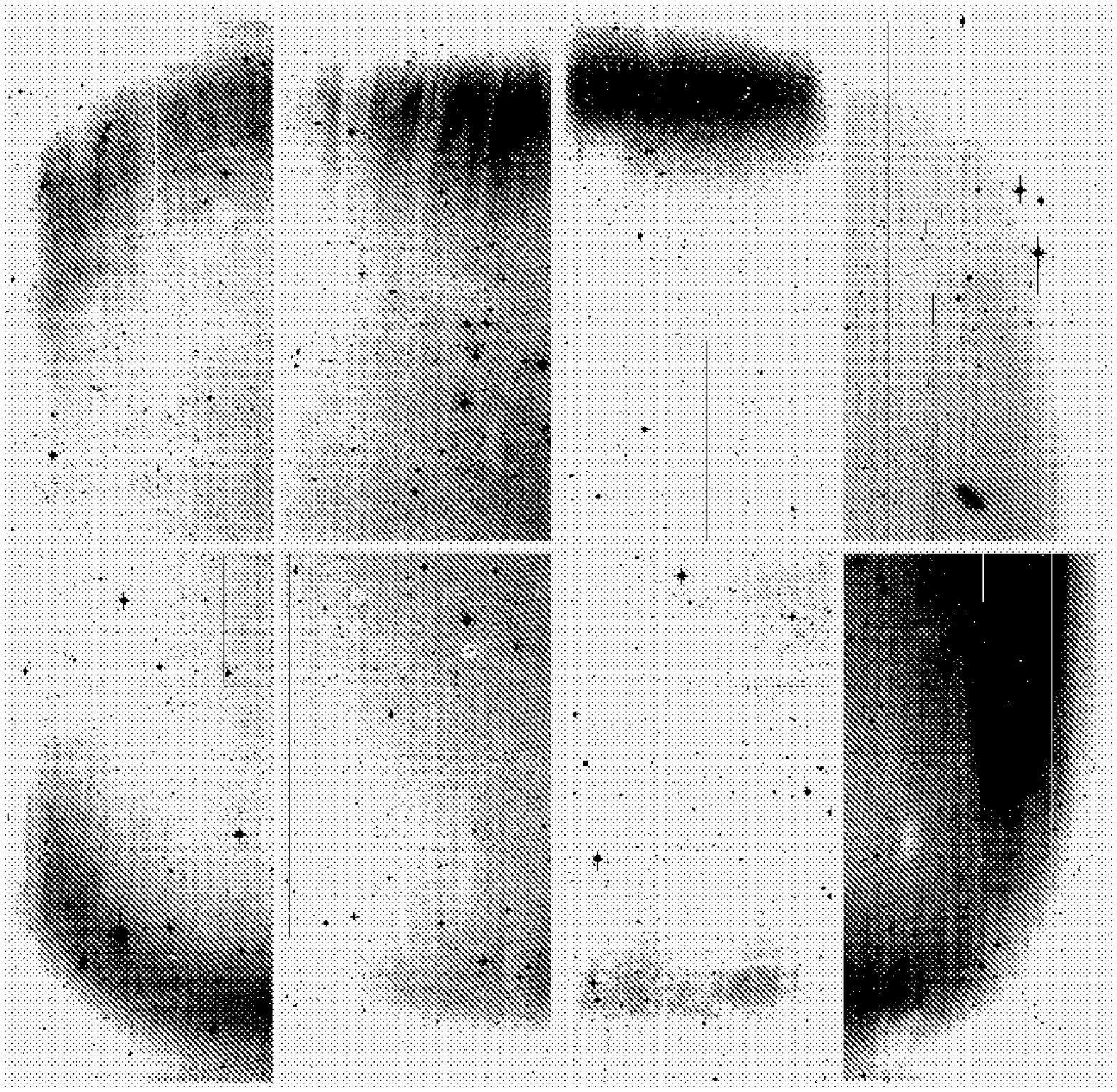}
\includegraphics{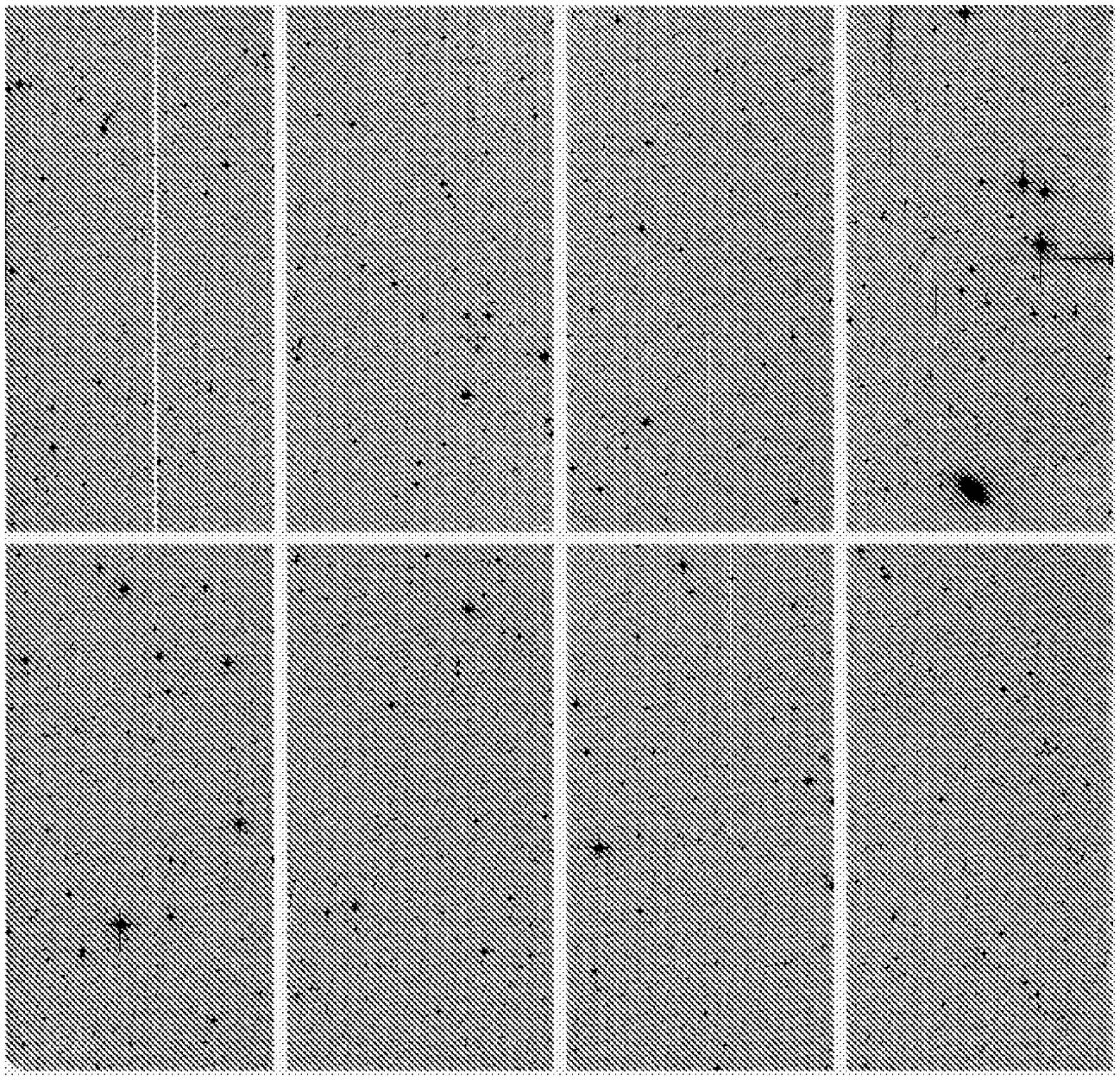}
\caption[]{Raw (left) vs. pre-reduced (right) OACDF image in the R band. The 
exposure time of this image is 900 seconds.}
\end{figure}

 Several twilight sky exposures were used to create a mean nightly
 twilight sky-flat. We define 
 $skyFlat= \langle skyFlats \rangle$. Rejection algorithms 
 to remove objects and cosmic particle heats were also applied. 
 
 The nightly master-flat in each filter was created using as much
 information as possible from the twilight sky-flats and the super-flats:
 the $supFlat / skyFlat$ ratio gives the information on the difference
 between the twilight sky flats and the average science frame
 and hence, it provides
 the pattern due to large-scale non-uniform illumination which should 
 be applied to flatten the science frames. The master-flat, 
 $Flat$, is then obtained combining the mean sky-flat with the super-flat 
 as follows:

\begin{equation}
Flat = smooth \Big( \frac{supFlat}{skyFlat} \Big) 
 \cdot skyFlat
\end{equation}

 In this way, the super-flat is used to correct the large scale 
 variations due to non-uniform illumination and the high S/N sky-flat 
 is used to correct the high-frequency pixel-to-pixel sensitivity 
 variations.

\subsection{Fringing correction}
 A fringing pattern was derived first using the ratio 
 $supFlat / skyFlat$ defined above. In case of 
 images with fringing such ratio is a superposition of the 
 fringing pattern, $frP$, itself and the  background.
 For images without fringing, such background is the same 
 as the ratio defined above.
 Therefore, in order to derive $frP$, it is sufficient to determine 
 the background of the $supFlat / skyFlat$ ratio and subtract it 
 from the ratio. Hence, the fringing pattern is given by:

\begin{equation}
frP  = \frac {supFlat}{skyFlat} - backgr\_fit \Big(\frac {supFlat}{skyFlat} \Big)
\end{equation}

 The background fit gives us both the fringing pattern, from eq.~2,
 and the information for the non-uniform illumination correction, 
 from the fit itself.
 Thus, the master-flat is determined using such fit in equation (1)
 instead of the smoothed ratio.

 Once the raw images are debiassed, the flat-field and  fringing 
 correction are applied as follows:
 
\begin{equation}  
PrI = \frac{Im}{Flat} -  \eta \cdot frP
\end{equation}

 where  $Im$ and $PrI$ are the debiased and pre-reduced image 
 respectively. The factor $\eta$ is a function of the sky background
 and the air-mass which varies from one
 scientific image to the other. It has been found that adopting  
 $\eta_n$ = bck$_n$ / $<$bck$>$,  where bck$_n$ is the mean background 
 of the n$^{th}$ image and $<$bck$>$ is the mean background of the 
 $supFlat/skyFlat$ ratio, gives good results. Typical values of $\eta$ run 
 from 0.8 to 1.2.
 A fringing pattern was obtained for every set of ditherings in every filter
 of every OACDF observing run.
 
\begin{figure}  
\vspace{6cm}
\includegraphics{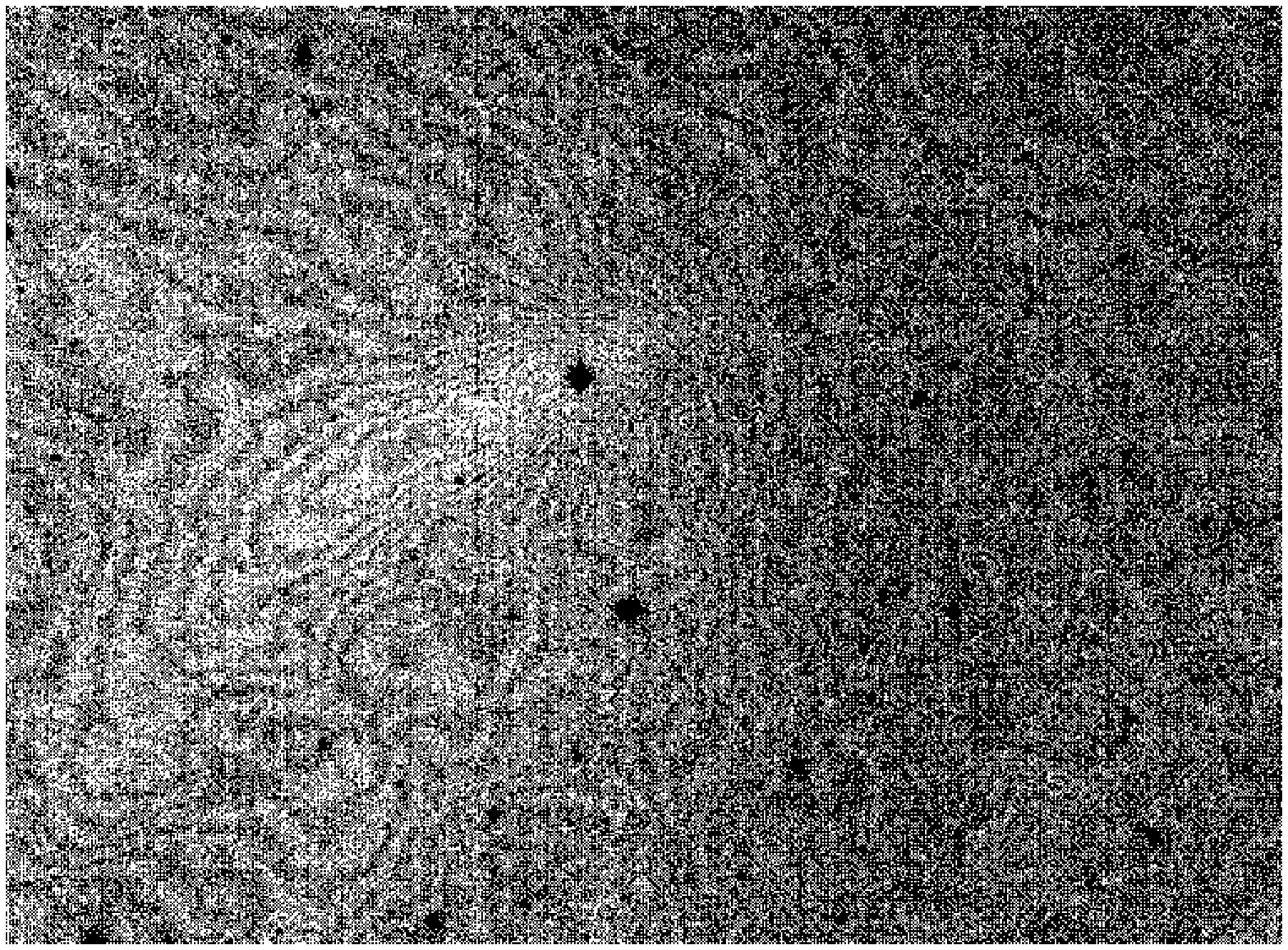}
\includegraphics{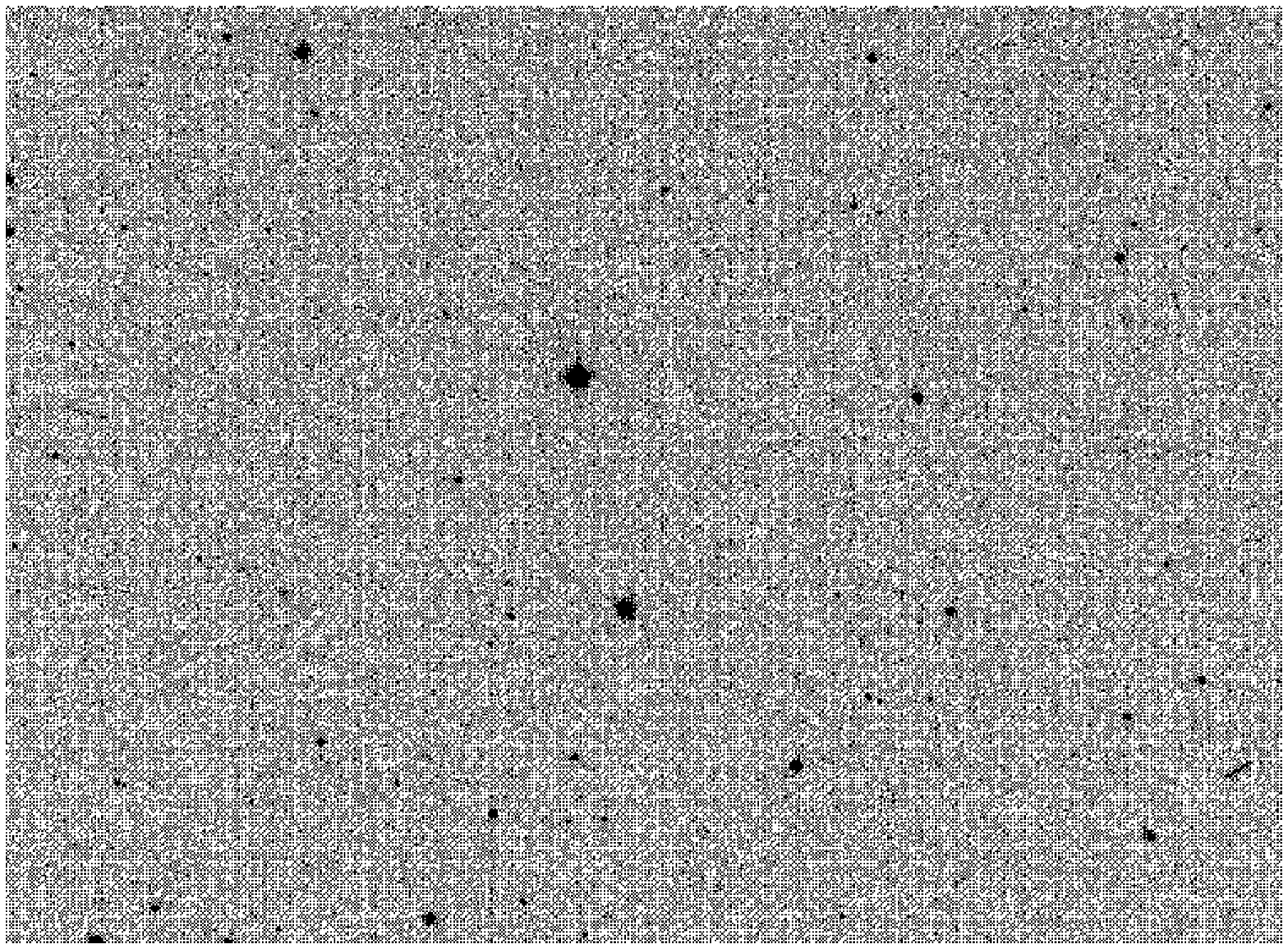}
\caption[]{Section of an image of the CCD No. 6 in the $\lambda 816$-band with
fringing (left). The final image, after correction by flatfield and fringing, 
is also shown (right).}
\end{figure}

Several tests were done in order to optimise the master
flat. The resulting images are uniform to better than 1\%.
In Figure~1, the comparison between a raw and pre-reduced image 
in the R-band is shown. In Figure~2, a section of an image of the 
CCD No. 6 of the mosaic is shown before and after the fringing 
correction (left and right panels respectively).

\subsection{The astrometry}
Two different strategies may be adopted for the astrometric calibration:
\begin{enumerate}
\item For each dithering sequence a reference exposure is chosen: an 
astrometric solution is computed for the CCDs in this exposure with reference 
to an external astrometric catalog. 
The astrometry for the other exposures is then computed by 
fitting the positions of sources to those measured in the reference frame.
\item For all CCDs the astrometric solution is computed simultaneously using 
both the external  astrometric catalog and the requirement that positions of 
sources  overlapping in different CCDs are the same ({\em global} astrometry).
\end{enumerate}

The first approach (see e.g. the {\it mscred} package in IRAF) works reasonably well 
in single pointings with compact dithers: however, it may give wrong results 
at the borders of the field, in particular if the offset between the exposures 
is large.
In the case of the OACDF the covered field consists of multiple overlapping
pointings and the offset between the exposures is typically $\ge$ 100 pixels. 
The global astrometry approach is therefore more convenient since it optimizes
the internal astrometric accuracy. The usage of the USNO A2 as the astrometric
catalog allows an RMS accuracy $\sim$ 0.3"; however the internal accuracy which may 
be obtained for overlapping sources in different bands is much lower. 
We therefore decided first to compute the astrometry
for the R-band image taking the USNO A2 as reference: we obtain an RMS accuracy of 
$\sim$ 0.4" (see Figure~3).  We then extracted a catalog of sources
from the finally coadded image and used it as the astrometric catalog for
the other bands. The RMS accuracy in this case is $\le$ 0.13" (c.f. Figure~3).

\begin{figure} 
\vspace{7cm}
\includegraphics{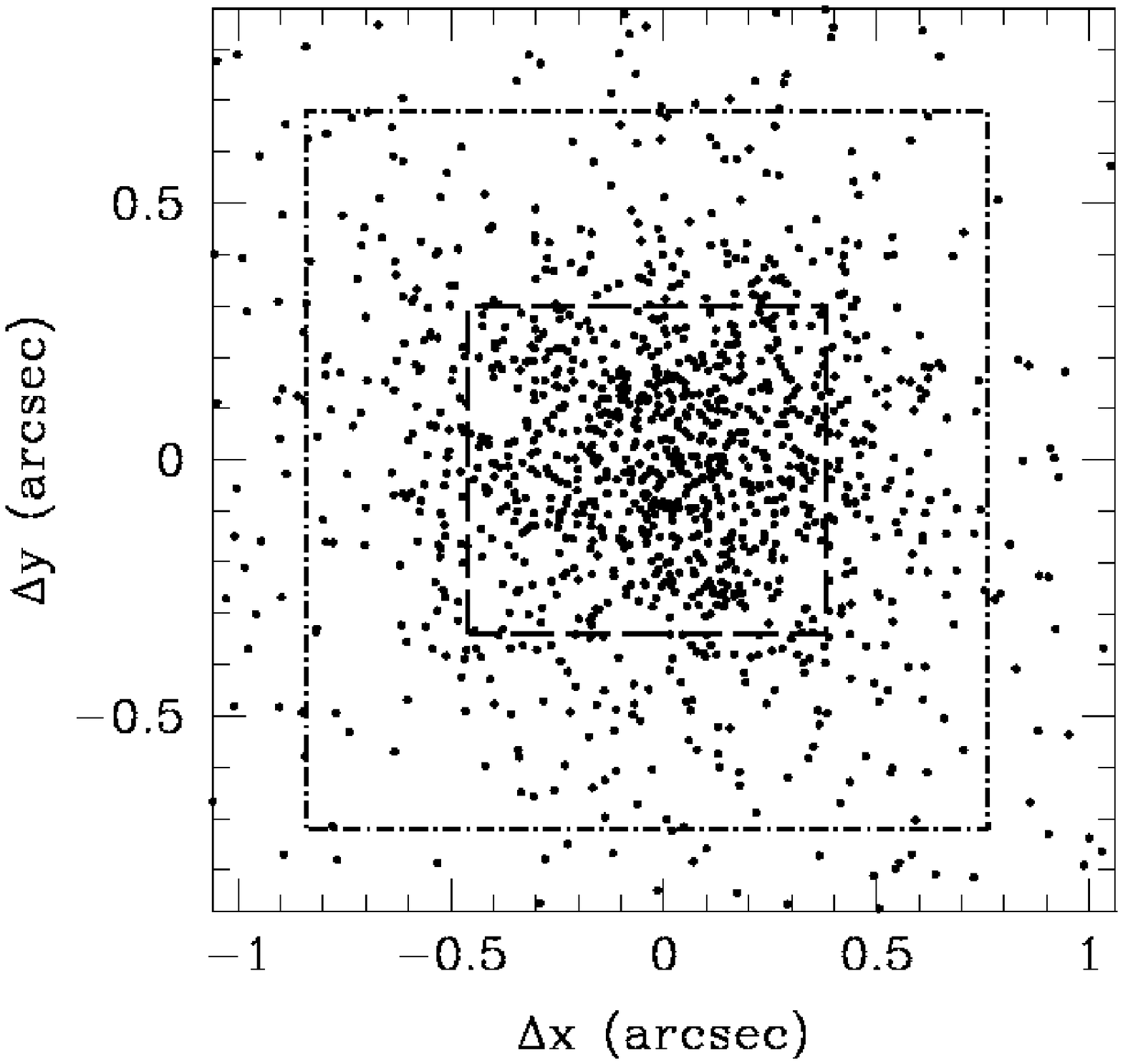}
\includegraphics{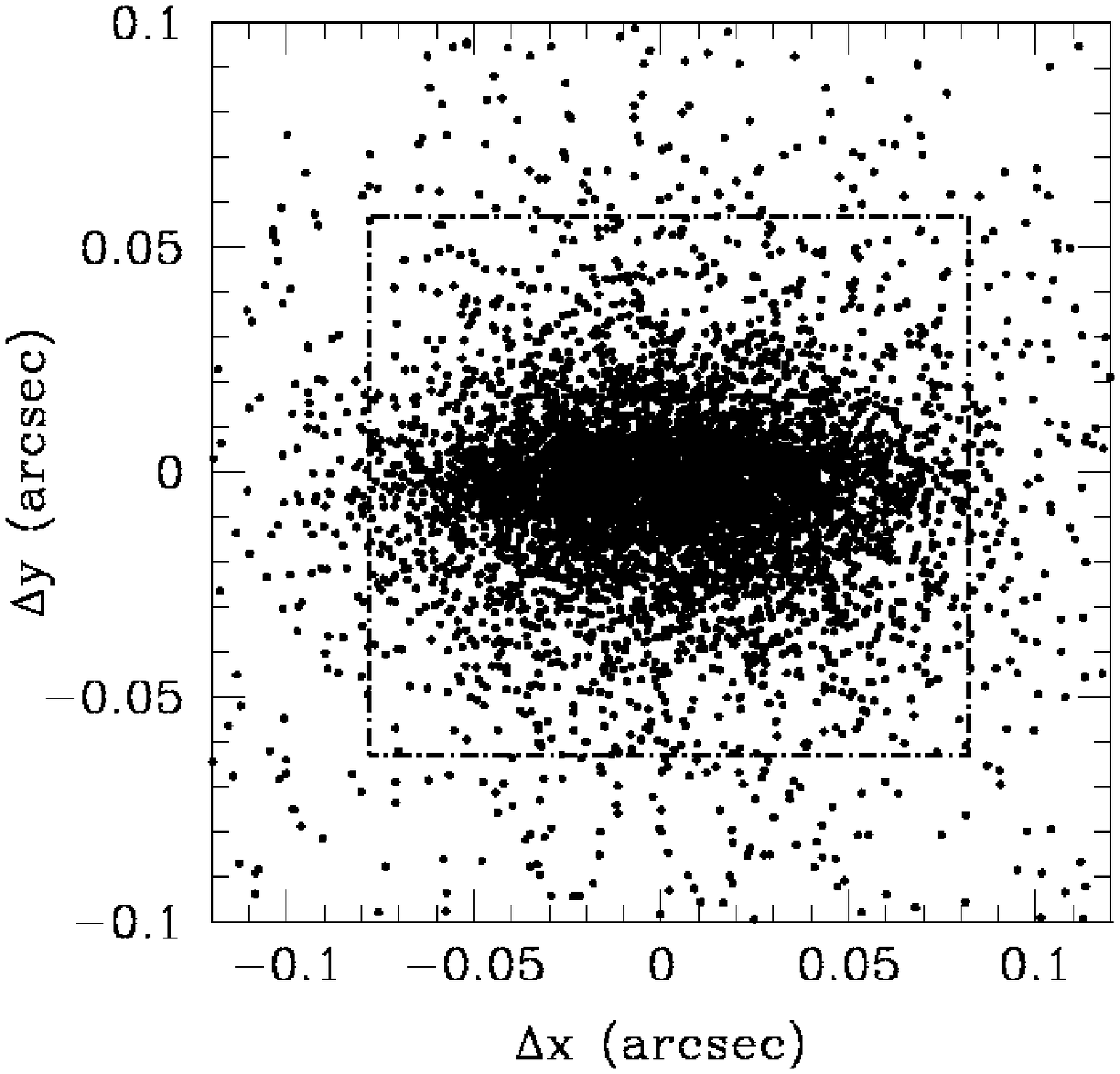}
\caption[]{Astrometric residuals. Left: R  vs. USNO; right: B vs. R. 
The dashed and dot-dashed lines show the residuals obtained for 68\% and
95\% of matched sources respectively.}
\end{figure}

The next step is to homogenize the internal photometry using an approach 
similar to that adopted for astrometry. The zero point for each frame is
written as: 
\begin{equation}
Z =  Z_{ph} + Z_r,  
\end{equation}
where $Z_{ph}$ is the photometric zero point computed from photometric 
standard fields (see later), $Z_{r}$ is an offset taking into account 
the changes in airmass, atmospheric conditions etc..
These offsets are computed requiring that overlapping sources in different 
exposures have the same flux. The photometry may be anchored to one or more 
'photometric' exposures for which $Z_r = 0$. 
 
Astrometry and photometry were computed using the {\em ASTROMETRIX} and
{\em PHOTOMETRIX} tools developed by M.R. in the framework of a 
cooperation of OAC with the TERAPIX team (IAP, Paris). Both are freely 
available in {\em http://www.na.astro.it/$\sim$radovich/wifix.htm}.

Finally, images were resampled, scaled in flux and coadded using the SWarp
tool developed by E. Bertin ({\em http://terapix.iap.fr/soft/swarp}).

\section{The photometric calibration}

\subsection{The broad-band calibration}
The B,V,R,I photometric calibrations were performed using the standard 
magnitudes and colours reported by Landolt (1992).
For the Landolt-98 field, we used the larger set of photometric 
standards by Stetson (2000). The transformation equations are as follows:

\begin{equation}
 B = b  - k_B \cdot X  + C_B \cdot (B - V)+  Z_B,
\end{equation}

\vspace{-0.7cm}
 
\begin{equation}
V = v  - k_V \cdot X  + C_V \cdot (B - V)+  Z_V,
\end{equation}

\vspace{-0.7cm} 

\begin{equation}
 R = r  - k_R \cdot X  + C_R \cdot (V - R)+  Z_R,
\end{equation}

\vspace{-0.7cm}

\begin{equation}
 I = ~i~ - k_I \cdot X  + C_I \cdot (V - I)+  Z_I,
\end{equation}
 
\noindent 
 where $X$ is the airmass, $b,v,r,i$ are the instrumental magnitudes, 
 $k_B,k_V,k_R,k_I$ the extinction coefficients, $C_B,C_V,C_R,C_I$ the 
 colour terms and $Z_B,Z_V,Z_R,Z_I$ the zero points in the B,V,R,I 
 bands respectively. The mean extinction coefficients for La Silla 
 were adopted.
 
 Some 100-200  standards were  used for the linear fits.
 The fit residuals are typically less than 2\%, but in some cases
 the residuals are as high as 5\% . The relatively high residuals 
 may be a consequence of the non-uniform illumination that will be 
 discussed in Section~4.3. 
 A few points with a particular high residual are due to stars falling
 partially in the CCD gaps and/or to photometric/astrometric errors
 in the standard catalogue. Such points were rejected by means of a sigma
 clipping algorithm.
 
 
 More details on the broad-band photometric calibration will be published
 elsewhere.


\subsection{The intermediate-band calibration}
 The flux calibration for the intermediate-band observations was performed
 using the observed spectrophotometric standard stars and following the 
 prescription by Jacoby et al. (1987).
  For a given astronomical source, the absolute flux $F_{\lambda}$
 measured at the earth in the band-width $\Delta \lambda$, is proportional
 to the extinction corrected count-rate measured in $\Delta \lambda$, and 
 inversely proportional to the peak of the filter transmission. 
 Let's call the proportionality factor as the $"counts-to-energy"$ conversion 
 factor $S_{\lambda}$. Thus,

\begin{equation}  
F_{\lambda} =  
\frac{ S_{\lambda} \cdot C_{\lambda} \cdot 
10^{ 0.4 \cdot k_{\lambda} \cdot X } }
{ T_{\lambda} }
\end{equation}

\noindent
where $C_{\lambda}$, $k_{\lambda}$, $X$ and
$T_{\lambda}$ are the count-rate, the extinction coefficient, the
airmass and the transmission peak in the band-width $\Delta \lambda$.
The $"counts-to-energy"$ conversion factor $S_{\lambda}$ is a measure
of the efficiency of the whole system (telescope + camera) and can be derived
from the observations of different spectrophotometric standard stars. 
Using $n$ standard stars, one can compute a mean 
$S_{\lambda} = \sum_{i=1}^n S_{\lambda}(i) / n$. 
The $S_{\lambda}(i)$ factor can be determined from the count-rate
$C_{\lambda}(i)$ of the $i^{th}$ standard star by  convolving
the spectral energy distribution, $F_{\lambda}(i)$, of the star 
with the filter transmission, $T_{\lambda}(i)$, using the following 
relation:

\begin{equation}  
 S_{\lambda}(i) = \frac{\int F_{\lambda}(i) \cdot  T_{\lambda} d\lambda}
{ C_{\lambda}(i) \cdot 
10^{ 0.4 \cdot k_{\lambda} \cdot X(i) } }
= \frac{F_{0}(i) \cdot \int T_{\lambda} d\lambda}
{ C_{\lambda}(i) \cdot 
10^{ 0.4 \cdot k_{\lambda} \cdot X(i) } }
\end{equation}

The last step in the above equation is justified by the fact that the
filters are narrow and the flux of the standard star can be considered 
as constant and equal to the value, $F_{0}$, in the center of the band.
Since the factor $\int T_{\lambda} d\lambda$ gives the equivalent width,
$W_{\lambda}$, of the filter, equation (10) can be written as follows:

\begin{equation}  
 S_{\lambda}(i) = 
\frac{F_{0}(i) \cdot  W_{\lambda} }{ C_{\lambda}(i) \cdot 
10^{ 0.4 \cdot k_{\lambda} \cdot X(i) } }
\end{equation}

Therefore, the $S_{\lambda}$ factor is determined on the basis
of the measured count-rate and the equivalent width of the filter.
Adopting the extintion coefficients $k_{\lambda} = k_I$ = 0.08
for all the intermediate bands, and measuring the count-rate 
$C_{\lambda}(i)$ of each standard by means of apperture photometry,
we obtained the $S_{\lambda}$ factors which are reported in Table~1
in units of 10$^{-16}~erg \cdot s^{-1} \cdot cm^{-2}  \cdot cts^{-1}$ 
Since the second observing run was partially non-photometric, the 
reported values are those for the first and third runs only.

\begin{table}[h]
\begin{center}
\caption{Mean $S_{\Delta \lambda}$ factors} 
\begin{tabular}{lccc}
\hline\hline
Filter        &     run1        &    run3	     \\
 (nm)         &                 & 		     \\
\hline
$\lambda$753  &  5.57$\pm$0.26  &   6.36$\pm$0.24   \\
$\lambda$770  &  6.20$\pm$0.34  &   7.14$\pm$0.20   \\
$\lambda$790  &  7.70$\pm$0.27  &   8.94$\pm$0.23   \\
$\lambda$815  &  6.60$\pm$0.29  &   7.69$\pm$0.35   \\
$\lambda$837  &  6.67$\pm$0.28  &   7.28$\pm$0.62   \\
$\lambda$914  & 11.67$\pm$0.85  &  12.77$\pm$0.67   \\
\hline
\end{tabular}
\end{center}
\end{table}

The AB magnitudes were determined from the relation:
$ 
m_{AB}(\nu) = -2.5 \cdot (log(F_{\nu}) + 19.436 )
= -2.5 \cdot log \Big( \frac{F_{\lambda} \cdot \lambda^2}{c} \Big) - 48.59, 
$
where $c$ is the speed of light. From equation (11) and after some algebra, 
it can be shown that:
$
m_{AB}(\lambda) = -2.5 \cdot log(C_{\lambda}) 
+ 2.5 \cdot log \Big( \frac{c \cdot W_{\lambda} }
{S_{\lambda} \cdot \lambda^2} \Big)
- k_{\lambda} \cdot X - 48.59 
= m_{AB}(\lambda) = m_{inst}(\lambda) 
 - k_{\lambda} \cdot X + Z_{\lambda}
$
, where $Z_{\lambda}$ is the zero point defined as 
$Z_{\lambda} = 2.5 \cdot log \Big( \frac{c \cdot W_{\lambda} }
{S_{\lambda} \cdot \lambda^2} \Big) - 48.59$.
In this way, the AB magnitudes  can be determined for a given 
OACDF observation with airmass $X$ adopting the extinction coefficient 
$k_{\lambda}$. The airmass to be used here is the one correspondent to the 
OACDF observation taken as reference dithering for the determination of the 
scaling factors as explained in Section~3.4. 


\begin{figure}  
\vspace{8cm}
\includegraphics{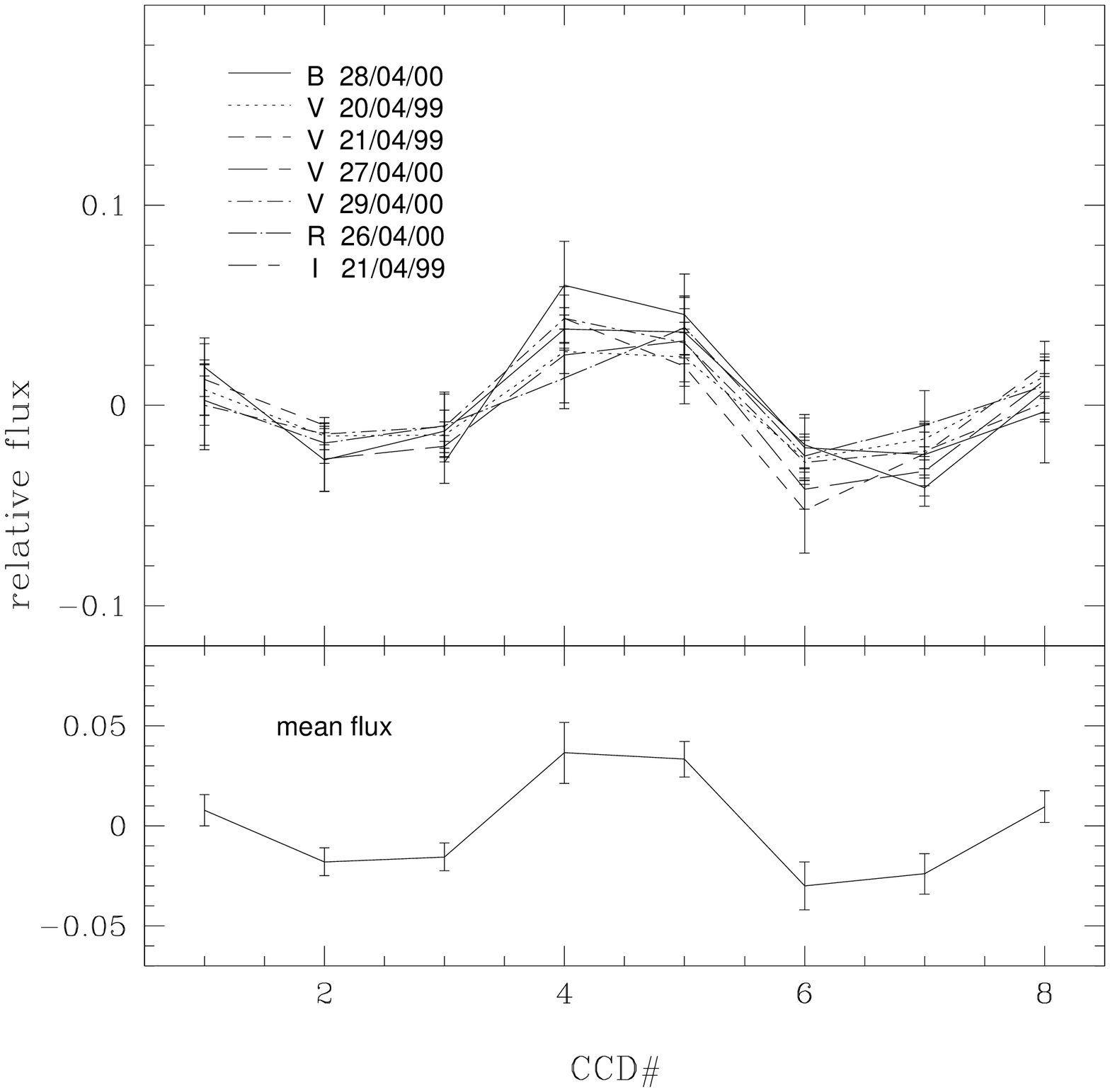}
\includegraphics{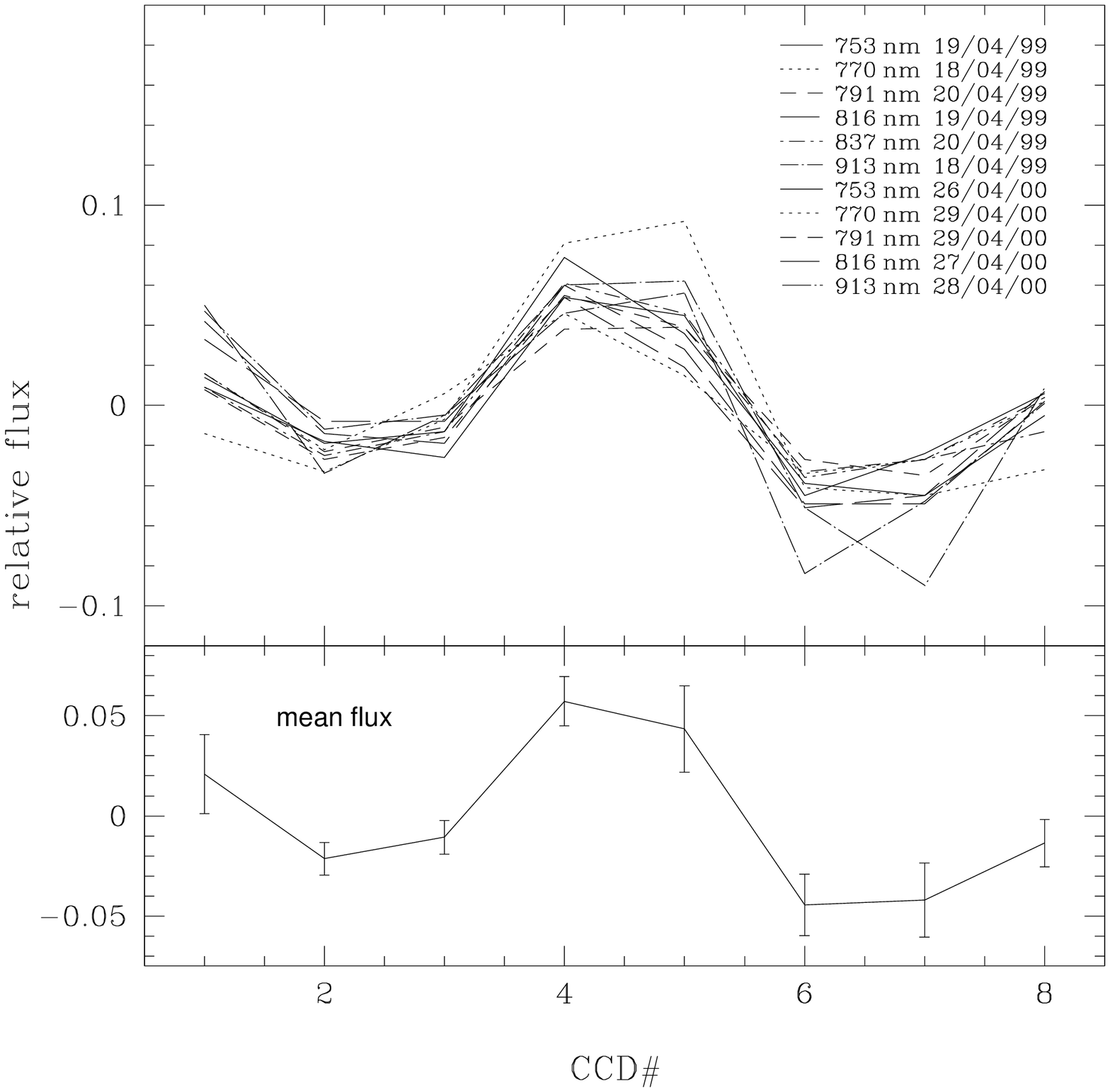}
\caption[]{The relative photometric offsets versus the CCDs for the broad-band 
(left) and medium-band filters (right) is shown in the upper panles. 
The average over the filters is shown in the lower panels.}
\end{figure}

\subsection{Inhomogeneity between different CCDs}
In order to test the accuracy of the photometric calibrations over
the whole mosaic, we observed a few selected standard fields in each
of the eight CCDs.

For the broad-band, we used the Landolt field PG1525-071, which 
contains five stars with (B-V) colours in the range from -0.198 to 1.109. 
In this way it has been possible to check the photometric homogeneity 
over the eight CCDs and its dependence from the star colour. These tests 
were repeated in each of the three runs using both MIDAS and IRAF aperture 
photometry. No significant differences of the results are found in the three 
observing runs. The results are summarised in Figure~4, where the relative 
photometric offset of each CCD with respect to the mean value over the 8 CCDs 
is shown. Within the errors, the trend is the same in the different broad 
bands (B, V, R, I) and different observing seasons (from April 1999 to 
April 2000).

A similar test was also done with the medium-band filters using the 
spectro-photometric standards Eg~274, Hill~600 and LTT~6248.
These stars were observed in all the three runs in each one of the eight
CCDs of the mosaic, in order to check whether there are differences in the 
$S_{\lambda}$ factor from one CCD to another. We note that the trend does 
not change significantly in the different runs (c.f. Fig.~4).
Note also that the photometric offset effect appears to be slightly stronger
in the medium-band filters with respect to the broad-band.

In conclusion, we find an offset in the flux of the stars which introduces 
an average uncertainty of $\pm$3\% in the broad-band B, V, R, I filters and
$\pm$5\% in the intermediate-band filters. This effect is quite constant in 
time.
 
The photometric offset, and hence the photometric inhomogeneity, may be 
attributed to an additional-light pattern caused by internal reflections 
off the telescope corrector. Such pattern, with an amplitude that depends
also on the exposure time, is present in every image, including flat-fields.
As such, it should be subtracted from every image before applying the 
flat-fielding correction, using an adequate scaling factor.

In order to be able to subtract the additional light, one must know the 
additional-light pattern. Unfortunately, for the WFI at the ESO 2.2m 
telescope such pattern is not well determined yet. The ESO 2.2m team is 
currently working on this issue.

\section{The catalog extraction}

As a consequence of the strategy adopted for the astrometry, the overlap between 
the same sources in different bands is $\le $0.1". 
Parameters were set in SWarp so that all the coadded images have the same 
number of pixels, scale and tangential point. It was therefore 
possible to combine the coadded images in all bands into one $\chi^2$ image
(see Szalay et al. 1999) that may be used for source detection. Actually, 
this is not straightforward since the depth of the survey in the 
medium-band filters is less than in the broad-band filters 
(see Table~2). Using only the $\chi^2$ image for the source detection 
would produce a large amount of spurious sources in the medium band filters
whose removal would not be trivial. Therefore, we produced a second set of 
catalogs for each band independently (single-image catalogs), in which the 
extraction parameters were optimized in order to obtain the maximum 
photometric deepness and the minimum number of spurious objects. 
An example of the results of this optimization are displayed in Figure~5, 
where the number counts in the B band, as well as the magnitude distribution
of all sources having a S/N $\simeq$ 10 are shown. The latter peaks at 
about B=24.6, which can be considered as the completeness limit of the B band 
catalogue.
As can be seen from Figure~5, the magnitude at which the number counts 
start declining is consistent with this value.

\begin{figure} 
\vspace{6cm}
\includegraphics{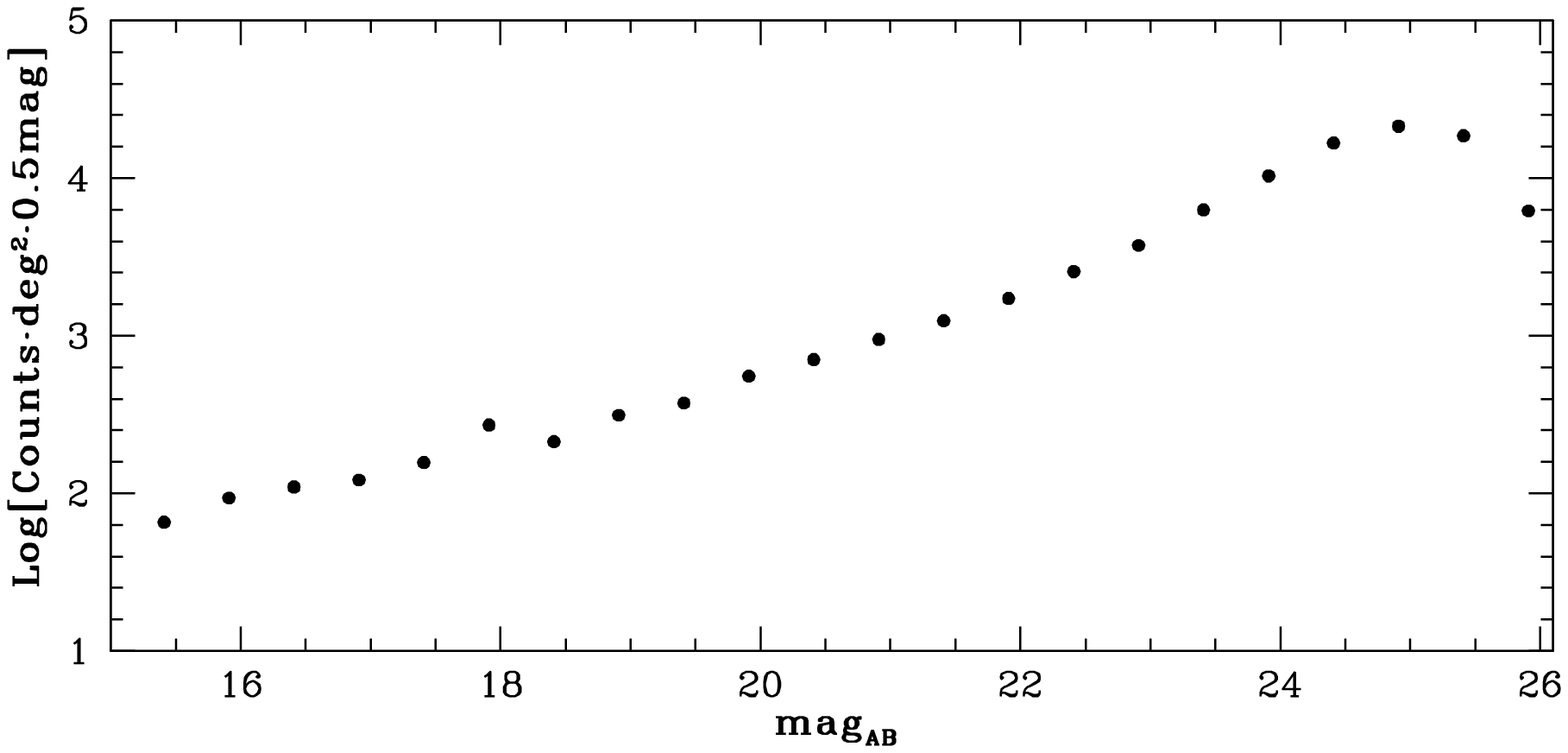}
\includegraphics{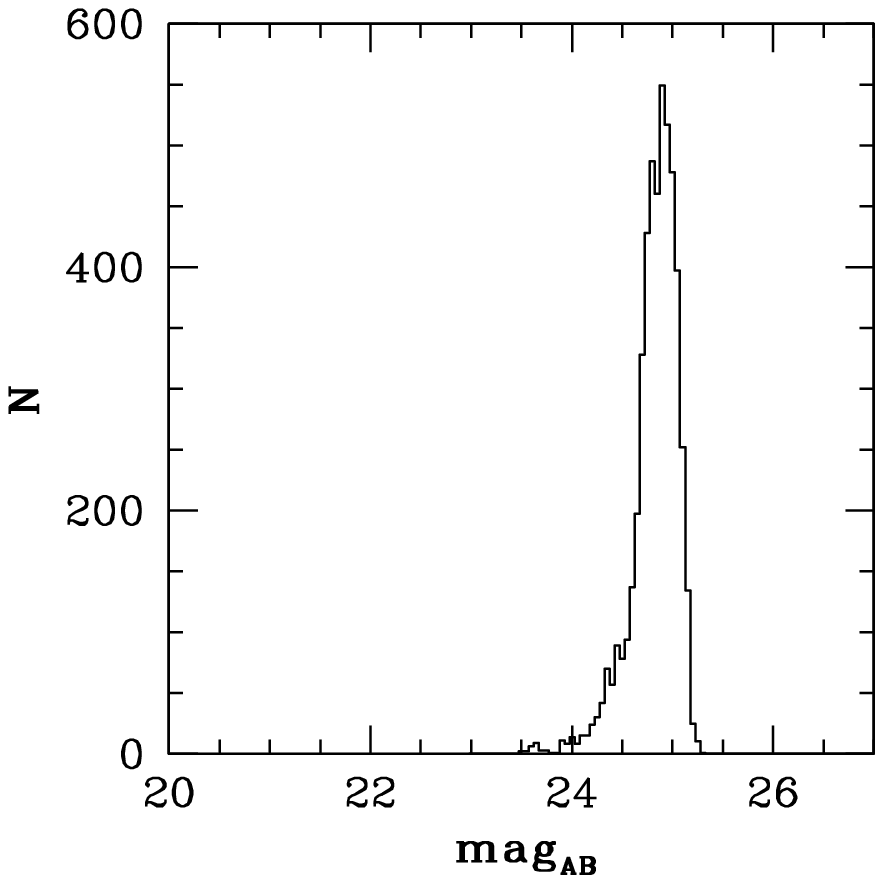}
\caption[]{The number counts of the sources detected in the B band
ins shown in the left panel, while the magnitude distribution of 
all the sources with S/N = 10 is shown in the right panel.}
\end{figure}

For each entry and each band in the catalog derived from the $\chi^2$ 
image, the photometry was kept only if that source was found in the 
single-band catalog. In this way we have been able to obtain, in one 
catalog, the full photometric OACDF information. The main advantage 
of this procedure is that the cataloge derived from the $\chi^2$ image
contains all the sources detected in at least one band, which allows, 
for example, to easily find drop-outs.

\begin{table}[h]
\begin{center}
\caption{Completeness magnitudes vs. wavelengths} 
\begin{tabular}{lccccccccc}
\hline\hline
S/N       &  B   &  V  &    R &   753 &	770 &  791 & 816 &  837 &  915\\
          &      &     &      &	      &	    &      &     &      &	\\
\hline
10  &  24.6  &   24.0  &   24.3 &  22.8 &   22.4 &   22.1 &   22.5 &   21.8 &  21.9	\\
~5  &  25.3  &   24.8  &   25.1 &  23.7 &   23.3 &   23.0 &   23.4 &   22.7 &  22.8	\\

\hline
\end{tabular}
\end{center}
\end{table}

A first check of the photometric quality of the catalogues has been done by 
the comparison of the colors of point-like sources with those expected from 
stars. To this aim, we took the Pickles' (1998) library of stellar spectra. 
The spectra were convolved with the WFI filter+CCD transmission curves. 
Point-like sources were selected using the CLASS\_STAR parameter in 
SExtractor. Figure~6 shows the good agreement between observed 
and simulated colors.

\begin{figure}[h]  
\vspace{8cm}
\includegraphics{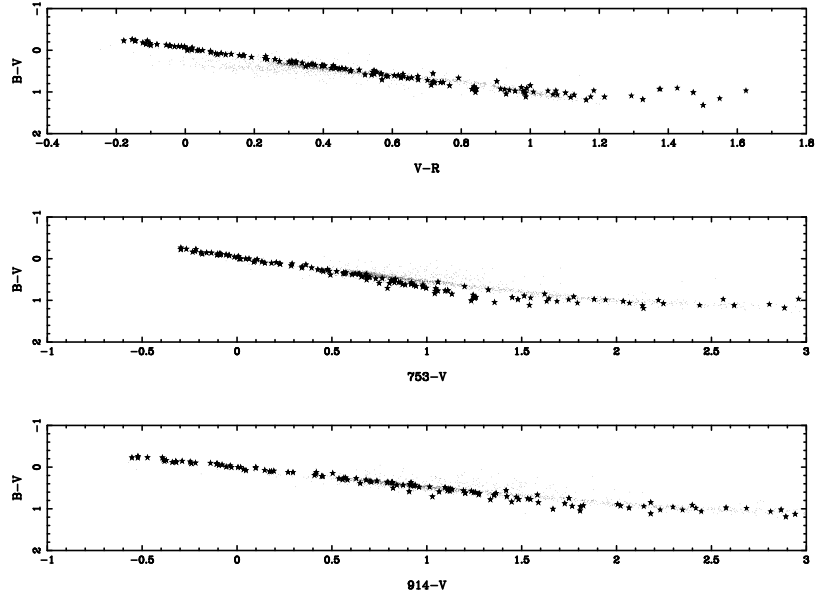}
\caption[]{The colors of point-like sources (small dots) are 
compared with those expected from the Pickles' library (stars).}
\end{figure}

\begin{figure}[h]  
\vspace{6cm}
\includegraphics{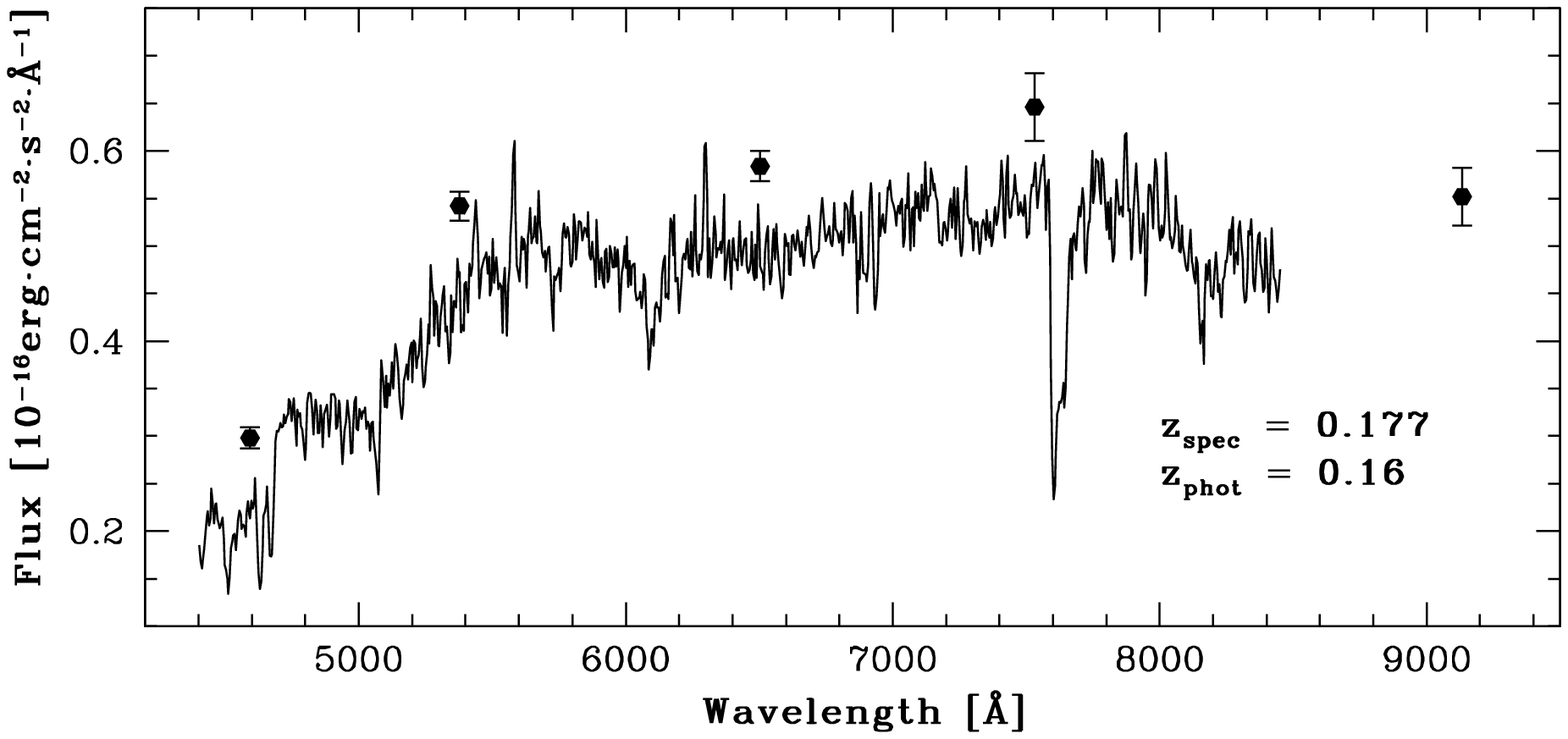}
\includegraphics{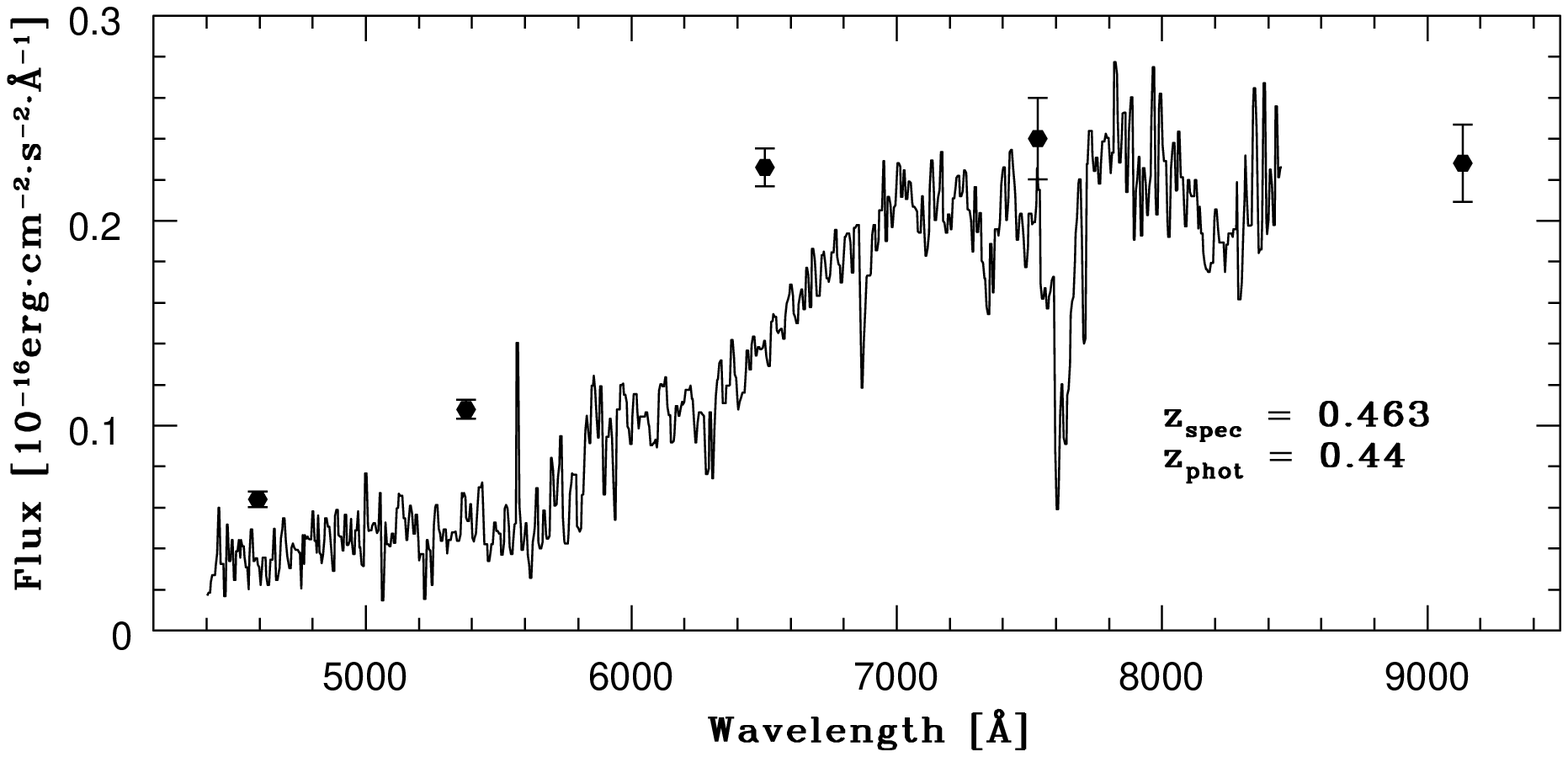}
\caption[]{Two examples of spectra of early type galaxies obtained during 
follow-up observations of the OACDF using EMMI-MOS at the ESO-NTT. The 
overplotted points represent the fluxes derived from the OACDF photometric 
data base. Such fluxes were used for the photometric redshift determination. 
The spectroscopic and photometric redshifts are also indicated. }
\end{figure}

\section{Some preliminary scientific results}

\subsection{A photomometric redshift catalogue}
Deep fields have become a favourite tool of observational cosmology, 
particularly in conjunction with the construction of multiwavelength 
datasets. As full spectroscopic coverage is usually impossible to 
obtain and the photometric redshift estimates have a certain 
degree of degeneracy, the only way to break this degeneracy is to have 
independent photometric information on a wider wavelength range, such 
as adding infrared colors or increasing the spectrophotometric resolution. 
The use of additional medium-band filters leads to a substantial gain in 
classification accuracy, as compared to broad-band photometry alone. 

Photometric redshifts for objects brighter than I$_{AB}$=22 were determined
using the HYPERZ package (see Bolzonella et al. 2000 for details). 
Spectroscopic redshifts, obtained from spectra of about ninety objects 
in the OACDF2 obtained during follow-up observations with EMMI-MOS 
at the ESO-NTT and that will be published in a forthcoming paper, were 
use to fine-tune the software parametrization.
The optimal set-up, which yield the most consistent results with the 
spectroscopic redshifts, was to use three broad-bands (B,V,R) and 
the intermediate-band filters centered at 753 and 914 nm. 
In Figure~7 we show some examples of spectra and their spectroscopic and 
photometric redshifts, when using such software set-up. The comparison 
of the spectroscopic and photometric redshifts yields an average residual 
equal to zero with a dispersion of 0.04. The complete catalogue with 
spectroscopic and photometric redshifts will be published in a 
forthcoming paper (Pannella et al., in preparation).

\begin{figure}[h]  
\vspace{10cm}
\includegraphics{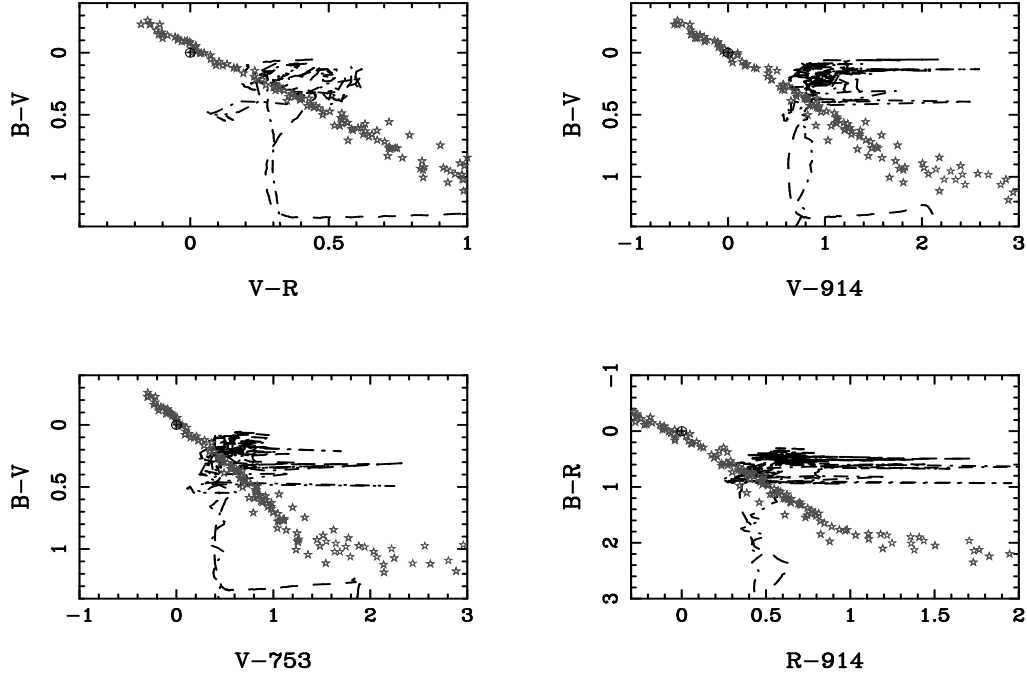}
\caption[]{Simulated colors of stars (from the Pickles' library) and active
galaxies (z=0-3); dashed line: AGN template (Vanden Berk et al. 2000); 
dot-dashed line: starburst template (Kinney et al. 1996). The spikes in 
the medium-band filters are produced by strong emission lines that enter 
the filter.}
\end{figure}
 
\begin{figure}[h]  
\vspace{15cm}
\includegraphics{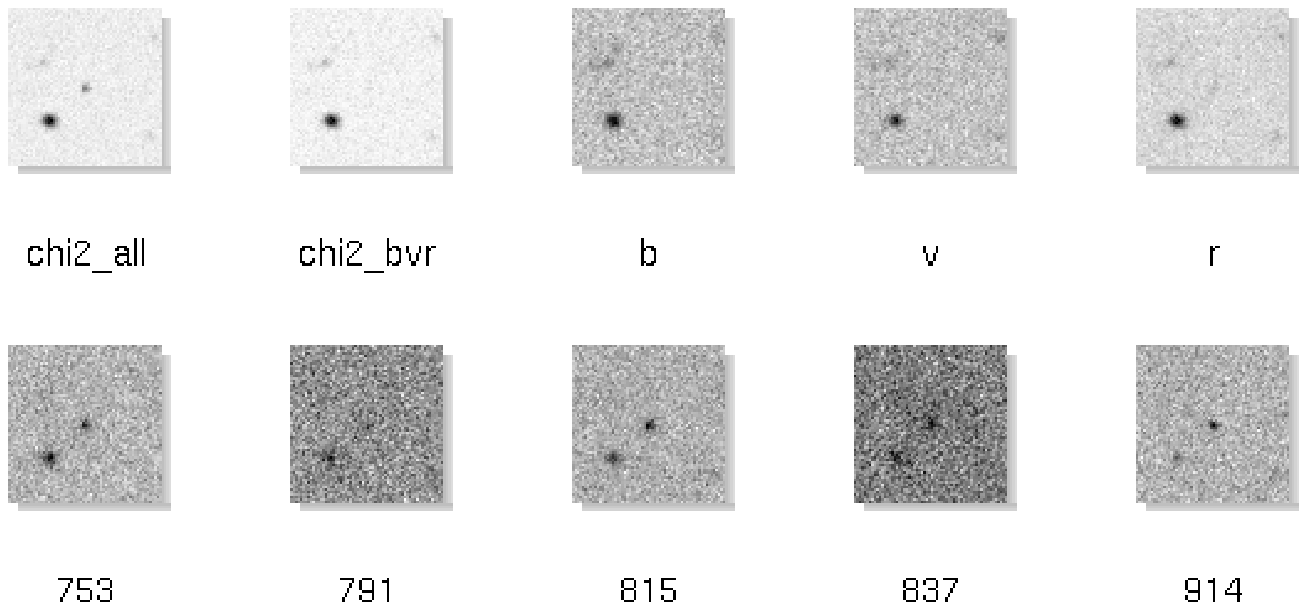}
\includegraphics{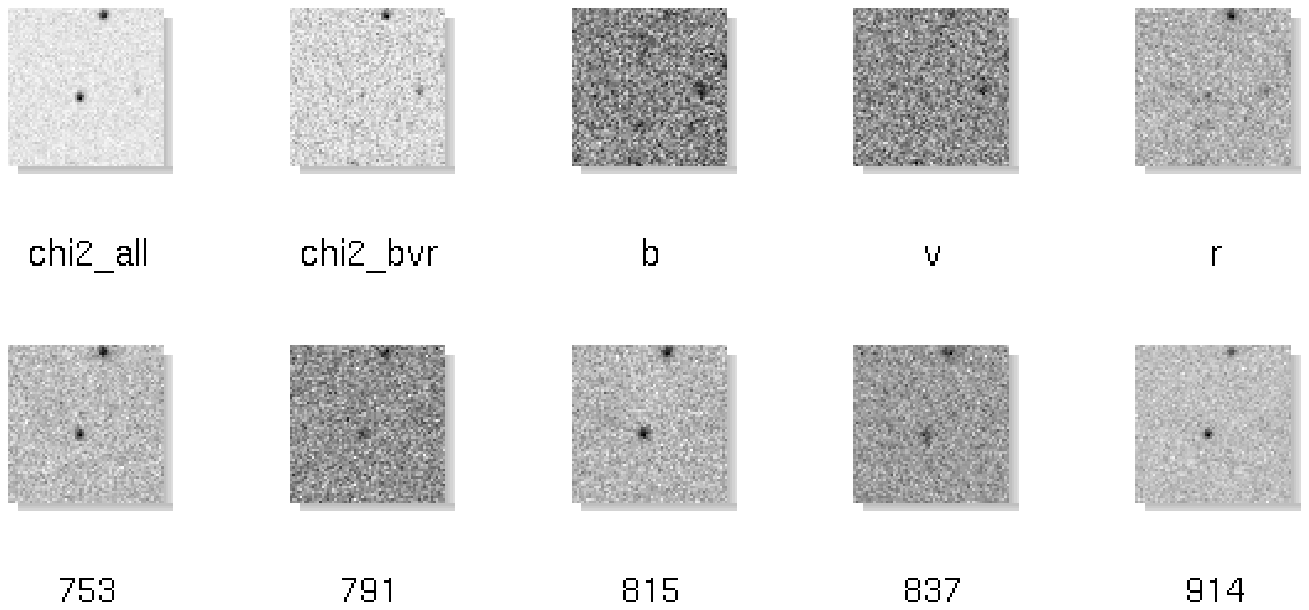}
\includegraphics{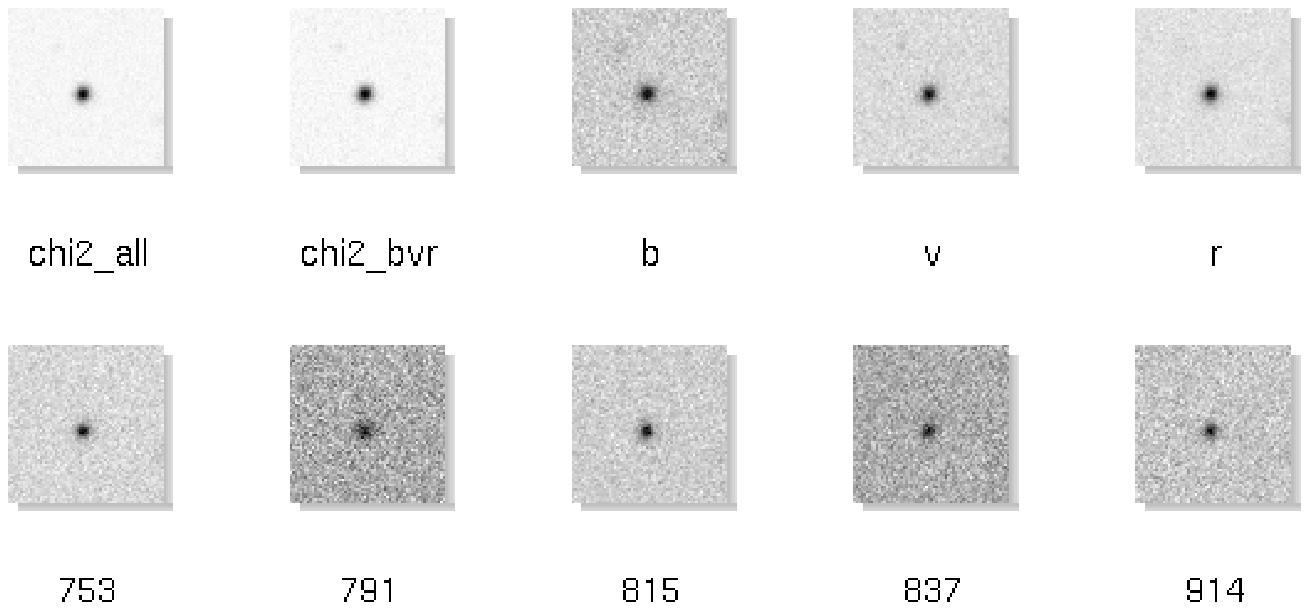}
\includegraphics{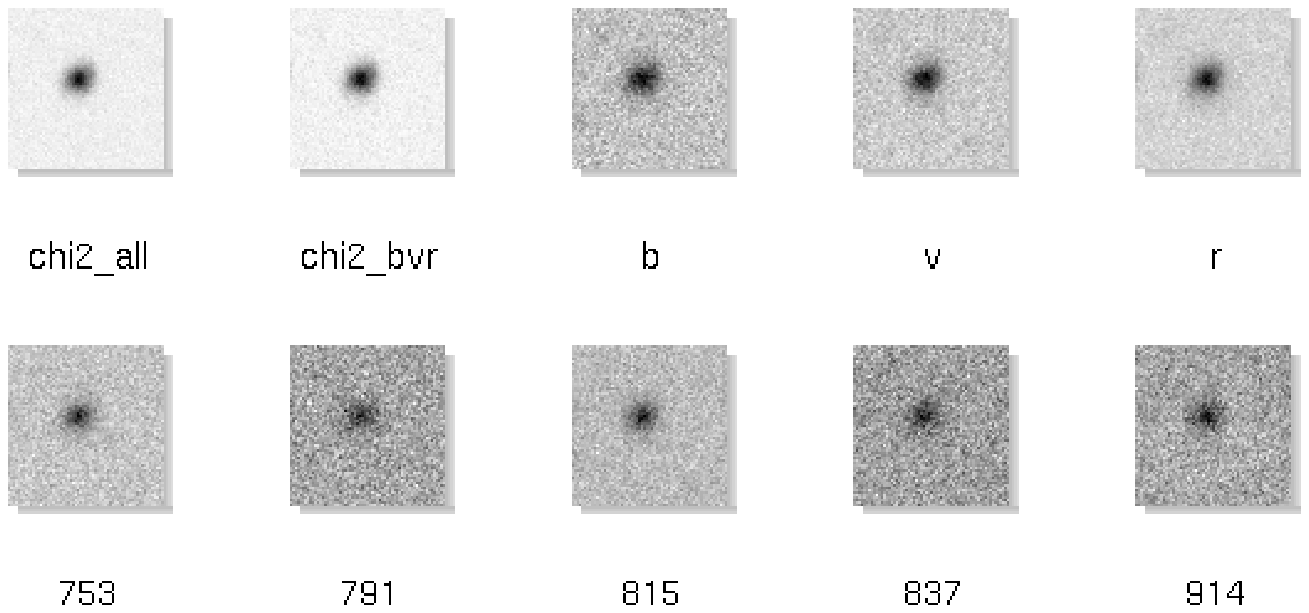}
\includegraphics{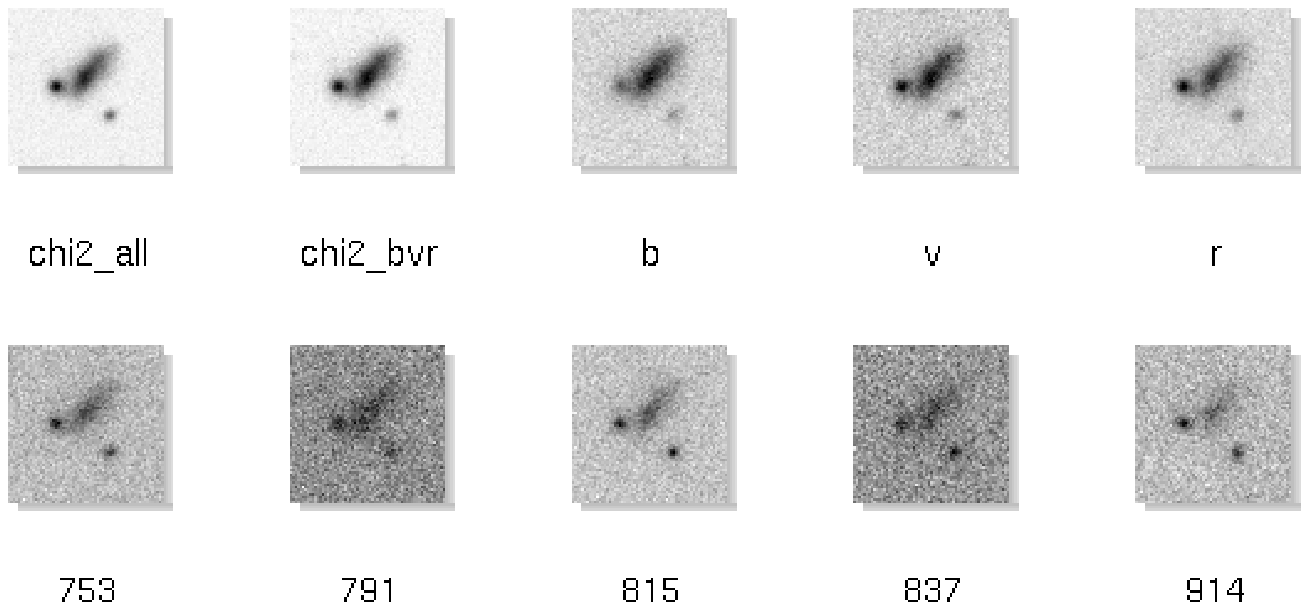}
\includegraphics{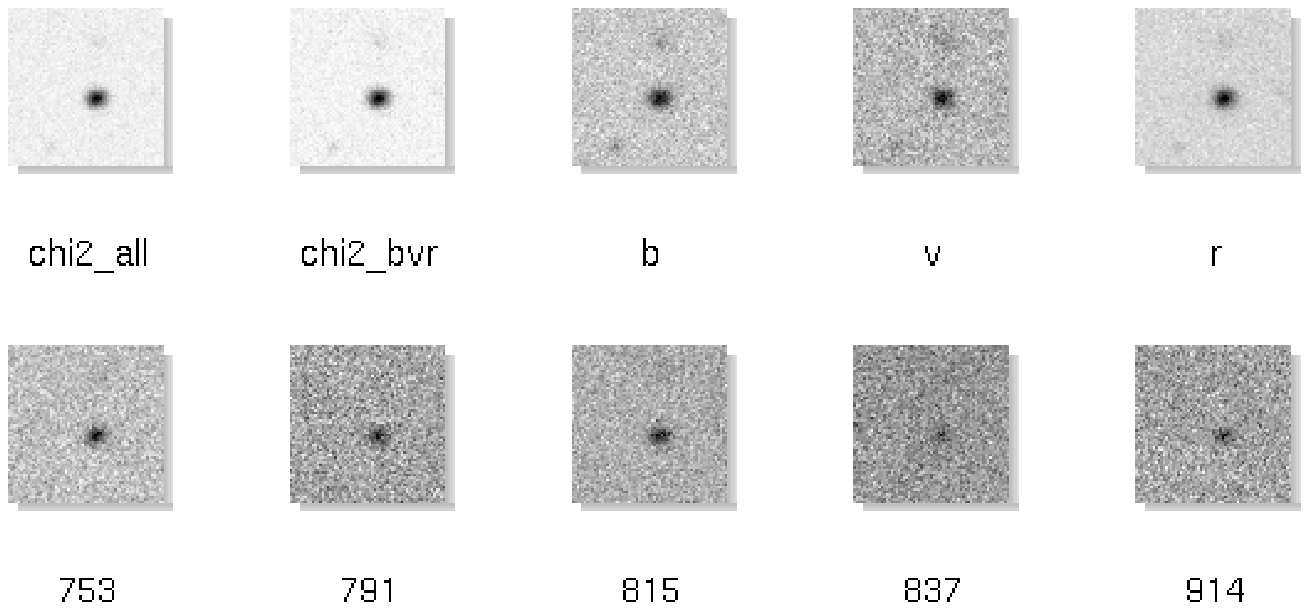}
\caption[]{Examples of candidates for: nearby active galaxies (top); 
high-z active galaxies (middle); R-dropouts (bottom). For each source, 
the images are taken from the $\chi^2$  built from all bands, the one built 
from the broad-band filters only, and each filter separately. chi2\_all
means the $\chi^2$ image for all the bands while chi2\_bvr means the $\chi^2$ image 
for the B,V and R bands.}
\end{figure}

\subsection{Search for AGNs and high-redshift objects}
We used HYPERZ  to compute the colors expected for our photometric system 
from normal and active (starburst + active galactic nuclei) galaxies. To 
this aim we used as templates the composite quasar spectrum from the 
Sloan Digital Sky Survey (Vanden Berk et al. 2000) and the starburst 
templates from Kinney et al. (1996) and convolved them with the filter+CCD 
transmission curves. The results are displayed in Figure~8.
This allowed to produce a first set of selection boxes for candidates of 
(i) nearby (z $<$ 2)  active galaxies (B-V $\le$ 0.2, 0.3 $\le$ V-R $\le$ 0.7)
and (ii) high-redshift (z $\ge 2$) active galaxies 
(B-V $\ge$ 0.8,  0 $\le$ V-R $\le$ 0.7). 
Thanks to the $\chi^2$ image, built from the images in all filters, it is 
also easy to look for (iii) sources detected in the medium-band filters 
only. These may be either emission-line galaxies where a line enters a 
medium-band filter, medium-redshift (z $\sim$ 1) elliptical galaxies 
where the 4000\AA\ break enters the R filter, or high-redshift (z $\ge 5$) 
galaxies  where the Ly$\alpha$ break enters the R filter, 
Some examples from these three groups are displayed in Figure~9.
Spectroscopic follow-up is being carried out, that should allow to 
validate and refine these criteria.

\subsection{Search for white dwarfs}

\begin{figure}[h]  
\vspace{7cm}
\includegraphics{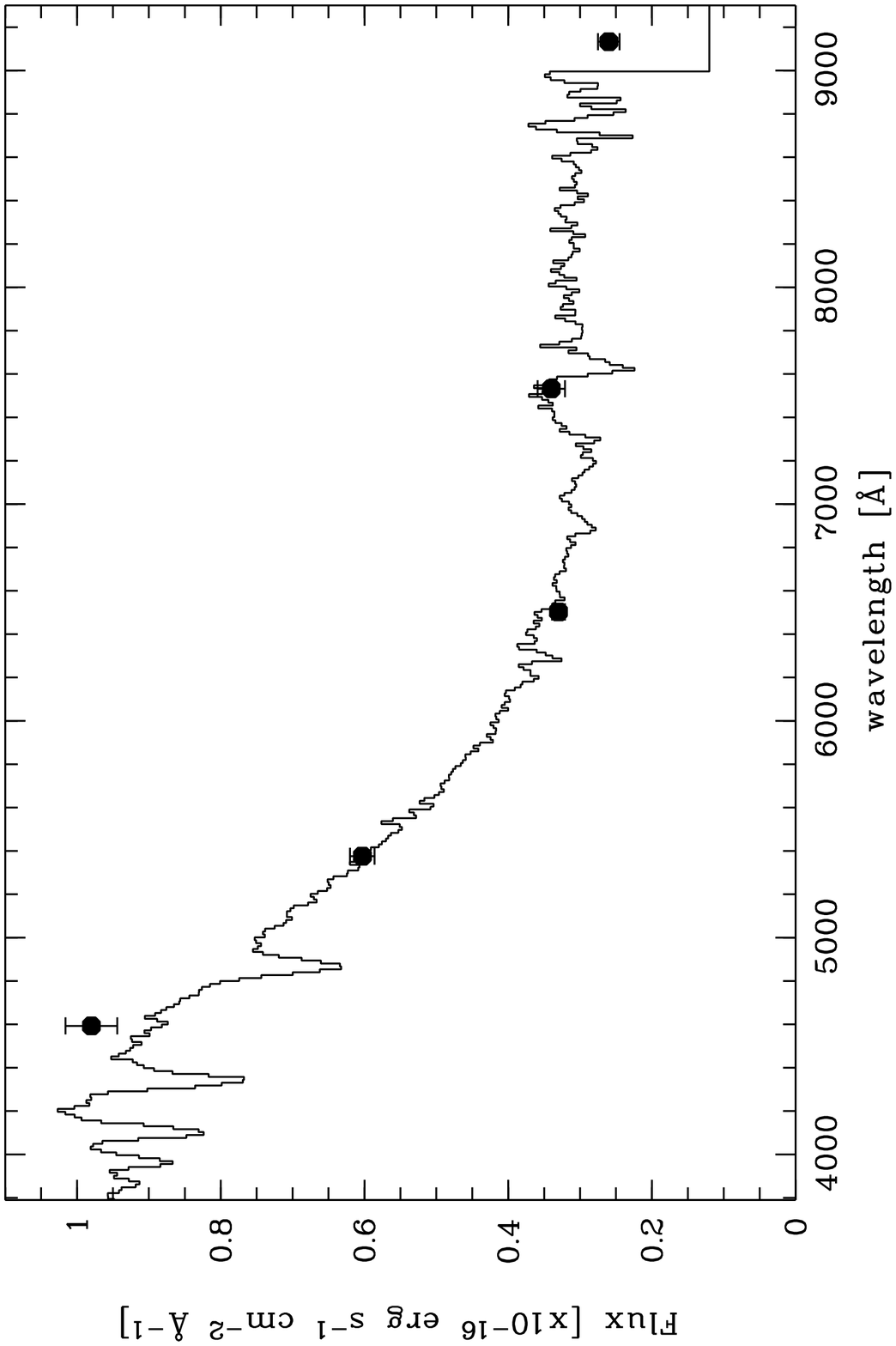}
\caption[]{Spectrum of a WD candidate obtained at the ESO 3.6m telescope with
EFOSC2. The fluxes obtained from aperture photometry are also reported (black
dots) and are in good agreement with the spectroscopic data.
The flat behaviour in the red part of the spectrum, starting near 7000 \AA,
suggests the presence of a cool companion.}
\end{figure}

The White Dwarf Luminosity Function (WDLF) contains crucial information on
the genesis of our galaxy: age of the galactic disk and halo, IMF, stellar
formation rate. Moreover the recent results of the MACHO+EROS microlensing
experiments indicate that the microlensing events are mainly produced by
halo objects with an average mass of $\sim$0.5 \msun (Alcock et al. 2000),
suggesting that part of the dark matter might be formed by halo white
dwarfs. The discovery of very cool WDs with \teff\,$<$\,4000\,K, whose 
statistic is almost totally unknown (less than 10 objects known up to date),
is fundamental to improve our knowledge on the questions addressed.

The aim of this research is to confirm spectroscopically the nature of a
few WD candidates with R magnitudes between 17.8 and 22.3, selected through
their colors from the OACDF survey.
For more details on the color selection see Silvotti et al. (2002).
Actually the WDs which can be well separated from the MS (Main Sequence) stars
are those with extreme (very high or very low) temperatures.
Moreover only the hydrogen WDs move out from the MS at very low temperatures 
(\lessim\,4000\,K) and become bluer in V--I because of the molecular hydrogen 
absorption (Hansen 1998), whereas the helium WDs become more and more red as 
they proceed along their cooling track, without leaving the MS region.

Presently spectra of about 10 WD candidates from the OACDF, obtained at the 
3.6m+EFOSC2 and NTT+EMMI ESO telescopes in March and April 2002, are under
reduction.
As an example, a spectrum of a White Dwarf is shown in Figure 10;
a preliminary comparison between the H$\beta$ FWHM and the LTE DA models from
Koester (private communication) gives an effective temperature of the order
of 14000\,K (considering a surface gravity $\log g$=8.0 in cgs units).

\section*{ACKNOWLEDGMENTS}
 We  thank E. Bertin for many useful discussions and suggestions regarding 
 catalogue extraction methods and F. Valdes for several suggestions on the 
 use of the IRAF mscred package.
 We thank A. Grado for his assistance during the observations with EFOSC2
 in March 2002. We are grateful to A. Di~Dato, M. Colandrea, K. Reardon and 
 M. Pavlov for their help with the INAF-OAC computers. We also thank the 
 ESO staff and the ESO 2.2m telescope team for their assistance during the 
 observations, and in particular E. Pompei for her inputs regarding the 
 ESO WFI characteristics. M.P. acknowledges financial support from the INAF-OAC.
 This project has been partially financed by the former Italian 
 {\it Ministero dell'Universit\`a e della Ricerca Scientifica e Tecnologica}
 (MURST).

\end{document}